\documentclass{PoS}

\title{Significance for signal changes in $\gamma$--ray astronomy}

\ShortTitle{Significance for Signal Changes}

\author{Dalibor Nosek\\ 
        Charles University, Faculty of Mathematics and Physics,
        Prague, Czech Republic\\
        E-mail: \email{nosek@ipnp.troja.mff.cuni.cz}}

\author{Stanislav Stefanik\\
        Charles University, Faculty of Mathematics and Physics,
        Prague, Czech Republic\\
        E-mail: \email{stefanik@ipnp.troja.mff.cuni.cz}}

\author{Jana Noskov\'a\\
        Czech Technical University, Faculty of Civil Engineering,
        Prague, Czech Republic\\
        E-mail: \email{noskova@mat.fsv.cvut.cz}}

\abstract{
We describe a straightforward modification of frequently invoked methods for 
the determination of the statistical significance of a  $\gamma$--ray signal 
observed in a counting process. 
A simple criterion is proposed to decide whether a set of measurements 
of the numbers of photons registered in the source and background regions 
is consistent with the assumption of a constant source activity. 
This method is particularly suitable for immediate evaluation of the stability 
of the observed  $\gamma\mbox{--ray}$ signal. 
It is independent of the exposure estimates, reducing thus the impact 
of systematic inaccuracies, and properly accounts for the fluctuations 
in the number of detected photons. 
The usefulness of the method is demonstrated on several examples. 
We discuss intensity changes for  $\gamma$--ray emitters detected 
at very high energies by the current  $\gamma$--ray telescopes 
(e.g. 1ES 0229+200, 1ES 1959+650 and PG 1553+113). 
Some of the measurements are quantified to be exceptional with large 
statistical significances.}

\FullConference{The 34th International Cosmic Ray Conference,\\
		30 July- 6 August, 2015\\
		The Hague, The Netherlands}

\usepackage{xcolor}

\newcommand\ApJ{ApJ }
\newcommand\AaA{A\&A }
\newcommand\EPJ{EPJ }
\newcommand\nima{NIM A }

\newcommand\beq{\begin{equation}}
\newcommand\beql[1]{\begin{equation} \label{#1}}
\newcommand\eeq{\end{equation}}
\newcommand\ben{\begin{eqnarray}}
\newcommand\een{\end{eqnarray}}
\newcommand\bea{\begin{array}}
\newcommand\eea{\end{array}}
\newcommand\bem{\begin{displaymath}}
\newcommand\eem{\end{displaymath}}

\newcommand\eqa[1]{Eq.~(\ref{#1})}

\newcommand\fig[1]{Fig.~\ref{#1}}
\newcommand\figs[1]{Figs.~\ref{#1}}
\newcommand\figg[1]{\ref{#1}}
\newcommand\tab[1]{Table~\ref{#1}}

\newcommand\sct[1]{Section~\ref{#1}}

\newcommand\qqc{\qquad \qquad}

\newcommand\wse{\vspace{-3.5cm}}

\newcommand\vf{0.85}

\newcommand\Ssum[2]{\sum \limits_{#1}^{#2}}

\newcommand\Pprod[2]{\prod \limits_{#1}^{#2}}

\newcommand\sgn{{\rm sgn}}

\newcommand\GeV{ \, {\rm GeV} }

\newcommand\LH{{\cal L}}
\newcommand\LHo{{\cal L}_{0}}
\newcommand\LHa{{\cal L}_{\rm A}}

\newcommand\nn{n }
\newcommand\non{n_{\rm on} }
\newcommand\noff{n_{\rm off} }
\newcommand\vnon{\vec n_{\rm on} }
\newcommand\vnoff{\vec n_{\rm off} }

\newcommand\nni{n^{i} }
\newcommand\noni{n^{i}_{\rm on} }
\newcommand\noffi{n^{i}_{\rm off} }

\newcommand\mon{\mu_{\rm on} }
\newcommand\moff{\mu_{\rm off} }
\newcommand\vmon{\vec \mu_{\rm on} }
\newcommand\vmoff{\vec \mu_{\rm off} }
\newcommand\hvmon{\hat{\vec \mu}_{\rm on} }
\newcommand\hvmoff{\hat{\vec \mu}_{\rm off} }

\newcommand\hvmono{\hat{\vec \mu}_{\rm on,0} }
\newcommand\hvmoffo{\hat{\vec \mu}_{\rm off,0} }

\newcommand\hmonoi{\hat{\mu}^{i}_{\rm on,0} }
\newcommand\hmoffoi{\hat{\mu}^{i}_{\rm off,0} }

\newcommand\moni{\mu^{i}_{\rm on} }
\newcommand\moffi{\mu^{i}_{\rm off} }

\newcommand\hmoffi{\hat{\mu}^{i}_{\rm off} }

\newcommand\valf{\vec \alpha }
\newcommand\alfi{\alpha^{i} }

\newcommand\Qi{Q_{i} }

\newcommand\hbeto{\hat{\beta}_{0} }

\newcommand\CL{{\rm CL} }

\newcommand\Noo{{\rm N} }

\newcommand\SLM{S_{\rm LM} }

\newcommand\PN{\Phi_{\rm N} }

\newcommand\pLM{p_{\rm LM} }

\newcommand\betr{\beta_{+} }
\newcommand\betl{\beta_{-} }

\newcommand\SR{S }
\newcommand\TR{T }
\newcommand\pR{p }
\newcommand\pC{p }

\begin{document}

\section{Introduction}
\label{Sec01}

The asymptotic Li--Ma technique~\cite{Lim01} has been traditionally 
used for more than thirty years in $\gamma$--astronomy in order 
to confirm positive results in searching for discrete $\gamma$--ray 
sources.
In high energy physics, similar likelihood--based techniques are often 
utilized to characterize the level of agreement between the data 
and the assumption of new phenomena~\cite{Cow01}.
Also different signal--to--background methods or more advanced techniques 
based, for example, on Bayesian reasoning are applied~\cite{Cou01}.
There is a large number of results based on the on--off analysis 
showing that a source producing events immersed in background 
is present in a suspected domain. 
However, quite different statistical tools are commonly applied 
in order to examine changes in a source rate.

In this study, the standard on--off methods~\cite{Lim01} 
are shown to be easily modified to test whether changes in 
the source activity are observed, while preserving all their 
important properties.
Our way of thinking is backed by the fact that the estimation 
of the source flux, which is usually based on complex calculations 
of exposure and spectral features, is not necessary when just changes 
in the source intensity are to be examined. 
We utilize the relationship between the source and background regions, 
which is well under control in current experiments, and deal  
with the source flux expressed in terms of the background flux.
Hence, based only on the numbers of recorded events in the regions 
of interest of known relative exposures, the proposed on--off
analysis is independent of most of the details of the detection.

The on--off method is applied to the available observational data. 
The recently collected very high energy (VHE) $\gamma$--ray data 
on the active galactic nuclei 1ES~1959+650~\cite{Ver02} is examined.
We analyze changes in the $\gamma$--ray data from the 1ES~0229+200 
source~\cite{Hes01,Ver01} and from the PG~1553+113 
blazar~\cite{Mag01,Hes02,Ver03}.

\section{The on--off method}
\label{S02}

We assume a set of $K$ independent 
observations with the numbers of events $\vnon = \{ \noni \}$ 
and $\vnoff = \{ \noffi \}$ detected in the on--source and 
off--source regions and with 
$\vec \nn = \vnon + \vnoff = \{ \nni \}$ in total, 
where $i=1,2,\dots,K$.
These on-- and off--source counts are assumed to come from the Poisson 
distributions with the parameters $\vmon = \{ \moni \}$ and 
$\vmoff = \{ \moffi \}$, respectively.
The relationship of the exposures of the corresponding regions 
is represented by a set of on--off parameters $\valf = \{ \alfi \}$.
The corresponding likelihood function given a set of $K \ge 1$ independent 
measurements of the numbers of events registered in the on-- and off--source 
regions is 
\beql{A01}
\LH(\vmon,\vmoff) = 
\Pprod{i=1}{K} P_{\noni}(\moni) P_{\noffi}(\moffi),
\eeq
where $P_{k}(a)$ is the Poisson probability to observe $k \ge 0$ 
events with the mean $a$, i.e. $k! P_{k}(a) = a^{k} e^{-a}$. 

\subsection{Test for intensity changes}
\label{S02a}

With the aim to test a constant rate of the source in a series 
of observations we examine whether the number of events registered 
in the on--source region when expressed in terms of background 
counts remains or does not remain constant. 
This hypothesis is represented by a set of $K$ conditions 
\beql{A02}
\moni = \alfi \beta \moffi, \qqc i=1,2,\dots,K,
\eeq 
where $\beta$ is an unknown source parameter.
For this, we consider the likelihood ratio 
$R_{K} = -2 \ln(\frac{\LHo}{\LHa})$. 
Here, $\LHa = \LHa(\hvmon, \hvmoff)$ and 
$\LHo = \LHo(\hvmono(\hbeto), \hvmoffo) = \LHo(\hbeto, \hvmoffo)$
denote, respectively, the likelihood function under the alternative
and under the null hypothesis. 
These functions are given by~\eqa{A01} and differ by the maximum likelihood 
estimates of the relevant parameters. 

Consider the case of $K \ge 2$ in which more than one pair of measurements 
are to be assessed whether they do not contradict to the hypothesis 
of a stable source.
In a standard way~\cite{Lim01,Cou01}, the on-- and off--source Poisson 
parameters $\hvmon = \vnon$ and $\hvmoff = \vnoff$ maximize the alternative 
likelihood function.
Under the null hypothesis, the maximum likelihood estimates 
follow from~\eqa{A01} where the constraints written 
in~\eqa{A02} are substituted, i.e.
\beql{A03}
\hmonoi = \alfi \hbeto \hmoffoi = \Qi(\hbeto) \nni, 
\eeq
where $\Qi(\beta) = \frac{\alfi \beta}{1 + \alfi \beta}$,
$\nni = \noni + \noffi$ is the total number of events collected in 
the $i$--th measurement and the maximum likelihood estimate of 
the source parameter $\beta$ is given implicitly by the equation
\beql{A04}
\Ssum{i=1}{K} \Qi(\hbeto) \nni = \Ssum{i=1}{K} \noni.
\eeq
If $\alpha = \alfi$, $i=1,2,\dots,K$, this equation reduces to an explicit 
expression $\alpha \hbeto \Ssum{i=1}{K} \noffi = \Ssum{i=1}{K} \noni$.

Using the maximum likelihood estimates of all relevant parameters,
the likelihood ratio of the on--off problem is $\TR = R_{K}(\hbeto)$ 
where 
\beql{A05}
R_{K}(\beta) = 2 \Ssum{i=1}{K}  
\left\{ \noni 
\ln \left[ \frac{1}{\Qi(\beta)} \frac{\noni}{\noni + \noffi} \right] + 
\noffi \ln \left[ 
\frac{\alfi \beta}{\Qi(\beta)} \frac{\noffi}{\noni + \noffi} \right] 
\right\}.
\eeq
Following Wilks' 
theorem~\cite{Wil01}, the statistic $\TR$ is asymptotically 
distributed as $\chi_{K-1}^{2}$ with $(K-1)$ degrees of freedom, 
i.e. $\TR \sim \chi_{K-1}^{2}$ as $\noni$ and $\noffi$ tend to infinity.
The $p$--value of the test, the probability that the assumption
of the constant ratio for all pairs of the on-- and off--source
parameters is rejected if it holds, is then given by 
$\pR = 1 - \Psi_{K-1}(\TR)$ where $\Psi_{K-1} = \Psi_{K-1}(x)$
is the cumulative distribution function of the $\chi_{K-1}^{2}$ statistic.
The significance with which the hypothesis of the constant 
source activity is rejected is given by the cumulative distribution 
function of a standard Gaussian variable, $\PN = \PN(x)$, i.e.
by the quantile function $\SR = \PN^{-1}(1 - \pR)$.

The above described procedure can be used when dealing with one 
pair of on-- and off-- source counts ($K = 1$). 
The source parameter $\beta$ is chosen in advance, from other 
measurements or theoretical considerations, for example.
The null hypothesis is $\mon = \alpha \beta \moff$ where $\alpha$ 
and $\beta$ are known parameters.
The likelihood ratio $\TR = R_{1}(\beta)$, is then asymptotically 
$\chi^{2}$ distributed with one degree of freedom, i.e. 
$\TR \sim \chi_{1}^{2}$ as $\non$ and $\noff$ tend to infinity.
The resultant test statistic may be written as
\beql{A06}
\SLM = s \sqrt{R_{1}(\beta)} = s \sqrt{2}  
\left\{ \noni 
\ln \left[ \frac{1}{Q(\beta)} \frac{\non}{\non + \noff} \right] + 
\noffi \ln \left[ 
\frac{\alpha \beta}{Q(\beta)} \frac{\noff}{\non + \noff} \right] 
\right\}^\frac{1}{2},
\eeq
where $Q(\beta) = \frac{\alpha \beta}{1 + \alpha \beta}$ and 
$s = \sgn(\non - \alpha \beta \noff)$. 
The statistic $\SLM$ can be assumed asymptotically as drawn from 
the standardized Gaussian distribution, $\SLM \sim \Noo(0,1)$.
The value of the sample variable $\SLM$ is interpreted 
as the asymptotic significance expressing that a $\mid \SLM \mid$ 
standard deviation event is observed above ($\SLM > 0$) or 
below ($\SLM < 0$) the given source intensity.
Then, the asymptotic $p$--value for an excess or deficit is 
$\pLM = 1 - \PN(\mid \SLM \mid)$.
Choosing $\beta = 1$, the hypothesis of no source 
($\mon = \alpha \moff$) is tested~\cite{Lim01}.
\begin{table}[ht!]
\begin{center}
{\footnotesize
\begin{tabular}{l l r r r r r r r}
\\\hline\hline\\[-2mm]
Source & Experiment & Period & $K$  & $\hbeto$ & $\TR$ & $p$--value & 
$S$ & $\langle \betl, \betr \rangle_{3\sigma}$ \\[2mm]
\hline\hline\\[-2mm]
1ES 1959+650 & VERITAS & 2012      &  3 & 4.89 & 173.5 & 
$9.35 \times 10^{-11}$ & 6.37 & \\[2mm] 
\hline\\[-2mm]
1ES 0229+200 & HESS    & 2005-2006 &  2 & 1.20 &   0.0 & 
0.960 &    0 & $\langle 1.11, 1.29 \rangle$ \\
             & VERITAS & 2009-2010 &  4 & 1.53 & 12.2 & 
$6.80 \times 10^{-3}$ & 2.47 & $\langle 1.46, 1.60 \rangle$ \\
             &         & 2010-2011 &  5 & 1.15 & 7.5 & 
0.110 & 1.23 & $\langle 1.02, 1.30 \rangle$ \\
             &         & 2011-2012 &  5 & 1.22 & 2.1 & 
0.716 &    0 & $\langle 0.94, 1.56 \rangle$ \\
             &         & 2009-2012 & 14 & 1.34 & 52.0 & 
$1.36 \times 10^{-6}$ & 4.69 & \\[2mm] 
\hline\\[-2mm]
PG 1553+113  & MAGIC   & 2005-2006 &  2 & 1.12 & 0.2  & 
0.673 &    0 & $\langle 1.07, 1.18 \rangle$ \\     
             & HESS    & 2005-2006 &  4 & 1.16 & 1.5  & 
0.693 &    0 & $\langle 1.10, 1.22 \rangle$ \\
             & VERITAS & 2010-2012 &  3 & 1.60 & 181.0 & 
$1.02 \times 10^{-10}$ & 6.36 & \\[2mm] 
\hline\hline\\[-2mm]
\end{tabular}
}
\end{center}
\caption{\small 
Results of on--off analysis.
The identification of objects, experiments and observational 
periods are listed in the first three columns.
The numbers of on-- and off--source pairs, maximum likelihood estimates 
of the source parameter ($\hbeto$), test statistics of the constant 
intensity hypothesis ($\TR$), corresponding $p$--values and 
significances ($S$) are summarized in the following five columns. 
The rightmost column contains, if possible, confidence intervals 
for the source parameter $\beta$ at a $3\sigma$ level.
}
\label{T01}
\end{table}

\subsection{Confidence intervals for intensity}
\label{S02b}

Suppose that the statistical test described in~\sct{S02a} failed 
to reject the constant source intensity at a given level of significance.
Then, one may construct a confidence interval of the source parameter 
at a higher level of confidence that covers its true value with 
the probability corresponding to this confidence level.
For this, we adopted the method of the profile likelihood ratio.
We assumed that the hypothesis given in~\eqa{A02} is satisfied and kept 
the source parameter $\beta$ as a tested parameter.
The likelihood ratio is then given by the likelihood function 
$\LHa = \LHa(\hvmon, \hvmoff)$ under the alternative and by 
the profile likelihood function given the source parameter $\beta$, 
i.e. $\LHo = \LHo(\hvmon(\beta), \hvmoff(\beta)) = 
\LHo(\beta, \hvmoff(\beta))$. 
In the alternative model, we used the maximum likelihood estimates 
of the on-- and off--source parameters, 
$\hvmon = \vnon$ and $\hvmoff = \vnoff$. 
The maximum likelihood estimates of the off--source parameters 
given $\beta$ are $\alfi \beta \hmoffi(\beta) = \Qi(\beta) \nni$.

Putting all these estimates into the likelihood ratio, we obtained
that a statistic $C(\beta) = R_{K}(\beta)$ follows asymptotically 
a $\chi^{2}$ distribution with $K$ degrees of freedom, 
i.e. $C(\beta) \sim \chi_{K}^{2}$. 
The confidence interval of the source parameter $\beta$ at a level 
of significance $\pC$ is constructed as a complement of 
a critical domain of the hypothesis test.
This way, $\beta \in \langle \betl, \betr \rangle$ at a confidence 
level $\CL = 1 - \pC$ when the quantile function of the $\chi_{K}^{2}$ 
distribution exceeds the statistic $C(\beta)$, 
i.e. $\Psi^{-1}_{K}(1 - \pC) > C(\beta)$.
Therefore, the endpoints of the confidence intervals are given by
\beql{A07}
C(\betl) = C(\betr) = \Psi^{-1}_{K}(1 - \pC).
\eeq
Note that the above introduced construction of the confidence interval
holds for any number of pairs of on-- and off--source counts, 
i.e. $K \ge 1$.
\begin{figure}[ht!]
\wse
\includegraphics*[width=\vf\linewidth]{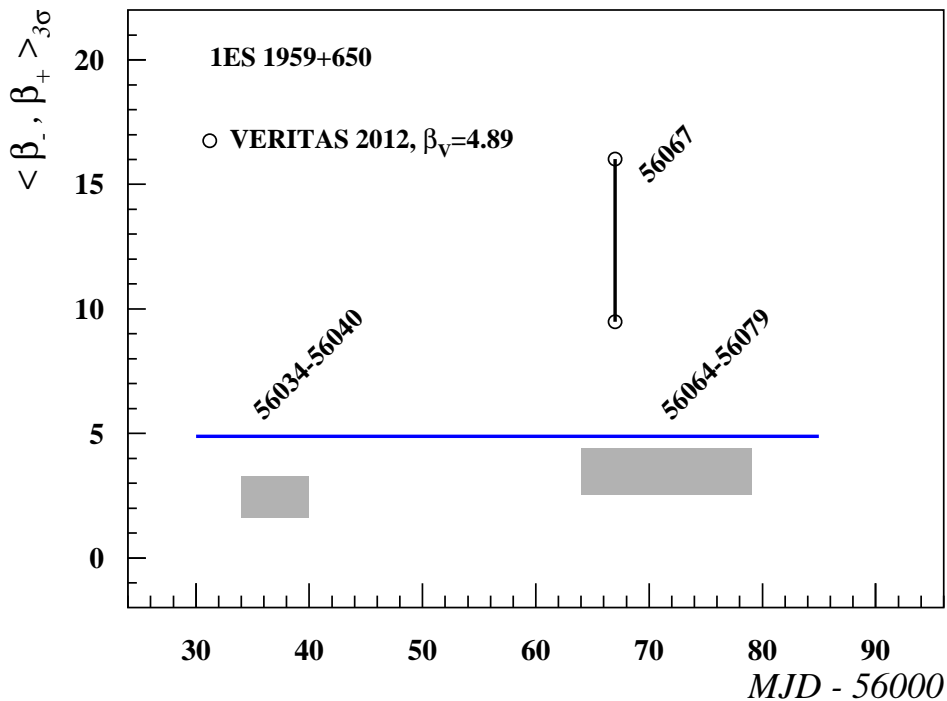}
\caption{\small
Confidence intervals at a $3 \sigma$ level of significance 
for the source parameter $\beta$ of 1ES~1959+650 are 
depicted as grey boxes. 
The exceptional measurement is shown as a line segment with circles.
Each of these limits is derived from a single on--off measurement.
Information about MJD of the detection is included~\cite{Ver02}.
The maximum likelihood estimate of the source parameter is 
visualized by a blue line. 
}
\label{F01}
\end{figure}
\begin{figure}[ht!]
\wse
\includegraphics*[width=\vf\linewidth]{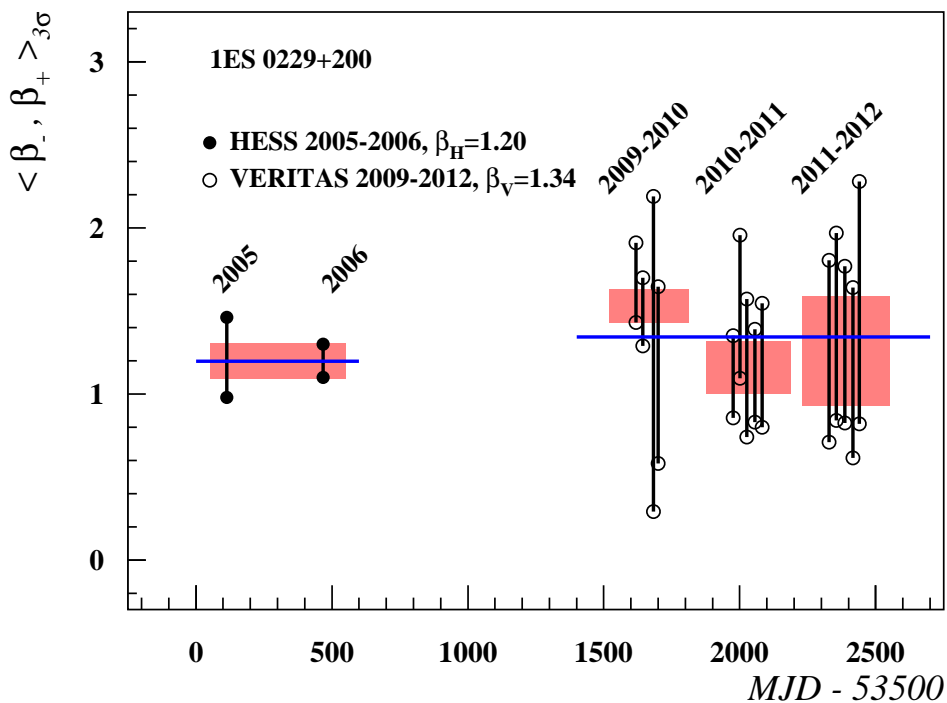}
\caption{\small
Confidence intervals at a $3 \sigma$ level of significance 
for the source parameter $\beta$ of 1ES~0229+200 are depicted.
Lines with full points and red box represent yearly measurements 
done by the H.E.S.S. instrument~\cite{Hes01}.
Lines with circles and red boxes refer to the monthly observations 
during the 2009--2012 VERITAS campaign~\cite{Ver01}.
Blue lines show the maximum likelihood estimates of the 
source parameters.
}
\label{F02}
\end{figure}
\begin{figure}[ht!]
\wse
\includegraphics*[width=\vf\linewidth]{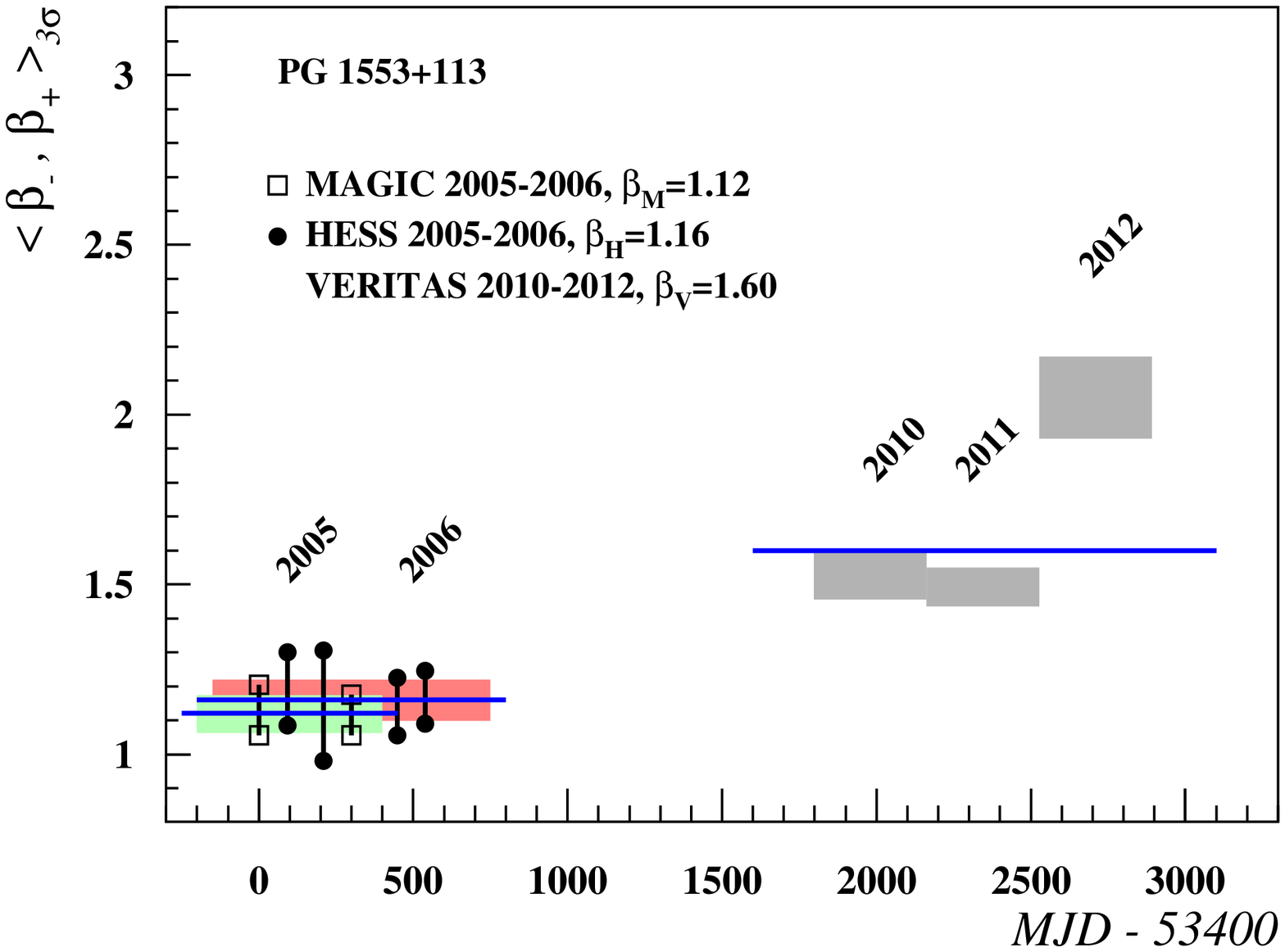}
\caption{\small
Confidence intervals at a $3 \sigma$ level of significance 
for the source parameter $\beta$ of PG~1553+113 are shown.
Green and red boxes are for estimates using the MAGIC and 
H.E.S.S. data~\cite{Mag01,Hes02}.
Grey boxes are for three individual on--off measurements reported by 
the VERITAS team~\cite{Ver03}. 
Blue lines correspond to the maximum likelihood estimates of 
the source parameters. 
}
\label{F03}
\end{figure}

\section{Examples}
\label{S03}

Within the proposed on--off method we examined VHE data for three objects. 
Specifically, we calculated test statistics $\TR$ as well as corresponding 
$p$--values and significances $S$ when testing intensity changes in 
a series of on--off measurements, see~\sct{S02a}.
If possible, we give confidence intervals for the ratio of the numbers 
of the on--source and background events, as described in~\sct{S02b}.
Our results are summarized in~\tab{T01} and in~\figs{F01}--\figg{F03}.

\subsection{1ES~1959+650}
\label{S03a}

The blazar 1ES~1959+650 was observed by the VERITAS setup between 
April 17, 2012 and June 1, 2012 (MJD 56034 and 56079)~\cite{Ver01}.
Typically, hundreds of VHE $\gamma$--ray events ($E > 315 \GeV$) were 
registered in each of the three reported observations.
A steady VHE flux of photons was rejected based on the light curve 
analysis with a very small $p$--value~\cite{Ver01}.

Results of our on--off analysis are shown in~\tab{T01} and in~\fig{F01}.
The on--off data taken from Ref.~\cite{Ver01} implies that 
the conjecture of a constant flux is rejected at a significance 
level of $\SR \approx 6.37$ ($p$--value $\approx 10^{-10}$).
For the three individual observations we also provide in~\fig{F01} 
confidence intervals for the source parameter $\beta$ at a $3\sigma$ 
level of confidence.
None of derived limits corresponds to the overall value of the source
parameter ($\hbeto \approx 4.89$) at a $3\sigma$ level of confidence.  
We learned that it suffices to judge the on--off data in order 
that changes of the source intensity were detected from the 1ES~1959+650 object.

\subsection{1ES~0229+200}
\label{S03b}

We employed the modified on--off scheme to investigate changes 
of the VHE $\gamma$--ray intensity associated with the BL Lac object 
1ES~0229+200.
The data used for this purpose comprise the results of the 2005--2006 
observations by the H.E.S.S. instrument~\cite{Hes01} as well as 
the 2009--2012 campaign of the VERITAS collaboration~\cite{Ver01}.
The numbers of registered on-- and off--source events in two yearly 
H.E.S.S. measurements ($E > 580 \GeV$) were above two hundred in 2005 
and even above one thousand in 2006. 
The three years campaign ($E > 300 \GeV$) conducted by the VERITAS 
experiment provided 14 month--long data sets with the numbers 
of events ranging from tens to several thousands.

Our on--off results are in agreement with the conclusion of the 
H.E.S.S.~collaboration~\cite{Hes01}. 
When examining these two measurements we found no support for 
the claim that the source intensity changed between 2005 and 2006, 
see~\tab{T01} and~\fig{F02}.

The evidence for variability present in the data sets has been 
advocated by the VERITAS team using the analysis of light
curve~\cite{Ver01}.
Our on--off results presented in~\tab{T01} allow us to soundly 
rule out that the source emission registered by the VERITAS setup 
is consistent with the long--term baseline intensity ($\hbeto = 1.34$) 
at a high level of significance, $\SR \approx 4.69$ 
($p$--value $\approx 10^{-6}$). 
The extraordinary nature of the 2009--2010 VERITAS observations 
is also documented in~\fig{F02} where the time sequences 
of the $\beta$--intervals are depicted. 
While most of the monthly data are consistent with background, 
the first monthly period in the 2009--2010 campaign is easily linked 
with a $3\sigma$ excess of measured counts above the three--years 
baseline activity.

\subsection{PG~1553+113}
\label{S03c}

The observations of BL Lac PG~1553+113 were performed by the MAGIC, 
H.E.S.S. as well as VERITAS instruments~\cite{Mag01,Hes02,Ver03}.
The light curve measured by the MAGIC telescopes in 2005 and 2006
($E > 200 \GeV$) showed no significant variations on a daily time 
scale~\cite{Mag01}.
Also, no evidence for the flux variation was detected during 
the 2 years of observations of the BL Lac PG~1553+113 by the H.E.S.S. 
setup ($E > 200 \GeV$) in the 2005--2006 campaigns~\cite{Hes02}.
The reported data set collected in the years 2010, 2011 and 2012 
by the VERITAS instrument~\cite{Ver03} contains three yearly 
measurements of thousands of detected events.
These VERITAS observations lead to the conclusion that the averaged 
flux detected in 2012 was increased by a factor 1.5 and 1.9 with
respect to the 2010 and 2011 observations, respectively.

In our on--off analysis presented in~\tab{T01} and in~\fig{F03}, 
we did not find any indication of intensity changes in the data 
collected by the MAGIC and H.E.S.S. telescopes.
On the other hand, when the on--off test is performed with 
the VERITAS data, a clear signature of an extraordinary 
observation registered in 2012 is returned.
This finding exhibits itself in a significance of $\SR \approx 6.36$
($p$--value $\approx 10^{-10}$) when the on--off test of the constant 
source activity is used.

\section{Conclusions}
\label{S04}

We studied changes of a signal in a chosen region when 
individual measurements of the number of source events are
accompanied by events due to background.
Based on the properties of the Poisson process, we introduced 
the asymptotic on--off test for the hypothesis that assumes 
a constant ratio of the source and background activity in 
the on--source region.
We described the construction of confidence intervals for 
the source intensity when expressed in terms of background.

The modified significance formulas were applied to the on--off data 
collected from three VHE $\gamma$--ray emitters.
We showed that, independently of the previous analysis, changes of 
the source activity are reliably quantified using on-- and off--source 
counts. 
The proposed on--off scheme is not only supported by convincing 
statistical motivation, but also relatively easy to use, yet 
sufficiently general.
Its independence of the complex estimation of detection conditions 
makes it suitable for exploring different types of intensity changes 
in different contexts.

\acknowledgments

This work was supported by the Czech Science Foundation under projects 
14-17501S and GACR~P103/12/G084.







\end{document}